\pdfoutput=1
\documentclass[10pt, aps, rmp,  twocolumn, superscriptaddress, floatfix, showkeys]{revtex4-2}
\usepackage[shortlabels]{enumitem}
\usepackage{multirow}
\usepackage{booktabs}
\usepackage{amsmath}
\usepackage{amsfonts}
\usepackage{graphicx}
\usepackage{calrsfs} 
\usepackage{times}
\usepackage[sort&compress]{natbib}
\usepackage{subfigure}
\usepackage{xcolor}
\usepackage{calrsfs}
\usepackage{soul}
\usepackage{booktabs, multirow}
\usepackage{comment}
\usepackage{verbatim}
\usepackage{framed}
\usepackage{lineno}

\usepackage{xcolor}
\usepackage{mdframed}

\newmdenv[
  backgroundcolor=gray!10,
  linecolor=gray!10,
  linewidth=0pt,
  innerleftmargin=10pt,
  innerrightmargin=10pt,
  innertopmargin=10pt,
  innerbottommargin=10pt,
  skipabove=10pt,
  skipbelow=10pt
]{graybox}

\providecommand{\keywords}[1]{\textbf{\textit{Keywords---}} #1}

\begin{document}
\newcommand{\fz}[1]{\textcolor{blue}{#1}}
\newcommand{\tbd}[1]{\textcolor{blue}{[TBD] #1}}


\title{Resolving Structural Uncertainty in Stochastic Dynamical Systems via Ensemble-Based Bayesian Inference}

\providecommand{\DFA}{Department of Physics and Astronomy ``Galileo Galilei'', University of Padova, Padova, Italy}
\providecommand{\PNC}{Padova Neuroscience Center, University of Padua, Via G. Orus 2, 35131 Padova, Italy}
\providecommand{\PCNM}{Padua Center for Network Medicine, University of Padua, Via F. Marzolo 8, 315126 Padova, Italy}
\providecommand{\INFN}{Istituto Nazionale di Fisica Nucleare, Sez. Padova, Italy}

\author{Francesco Zambelli}
\email{francesco.zambelli@unipd.it}
\affiliation{\DFA}
\author{Valeria D'Andrea}
\affiliation{\DFA}
\author{Manlio De Domenico}
\email{manlio.dedomenico@unipd.it}
\affiliation{\DFA}
\affiliation{\PNC}
\affiliation{\PCNM}
\affiliation{\INFN}

\vspace{0.4 cm}
\begin{abstract}
\vspace{0.2 cm}  
Recovering the interaction structure of complex stochastic dynamical systems from sparse time-series observations is an inherently ill-posed inverse problem. The posterior over network configurations is high-dimensional, highly degenerate, and shaped by the non-equilibrium nature of the available data. To address this, we propose a fully Bayesian, ensemble-based inference framework. Our approach grounds a continuous-time Stochastic Differential Equation (SDE) likelihood formulation---acting mathematically as an Onsager-Machlup quasi-potential for non-equilibrium paths---within a non-parametric Minimum Description Length (MDL) structural prior. This framework rigorously penalizes model complexity and avoids kinetic overfitting by forcing continuous parameters into a finite vocabulary of discrete categories. The joint posterior is optimized using a block-coordinate scheme combining sub-quadratic Markov Chain Monte Carlo (MCMC) topological sampling with deterministic Nelder-Mead parameter optimization. Using highly non-linear macroscopic population dynamics as a benchmark, we demonstrate that inference requires non-stationary transient data to resolve causal information. Furthermore, we prove that propagating structural uncertainty via ensemble-based posterior analysis produces significantly more robust and reliable predictions for downstream perturbation outcomes than traditional single-point (MAP) estimates. The framework is highly generalizable, capturing fundamental topological constraints even under severe generative model mismatches, offering a robust tool for uncertainty-aware inference in complex stochastic systems.
\vspace{0.2 cm}
\end{abstract}


\maketitle

\section{Introduction}

Inferring the structure of a complex dynamical system from time-series observations is a fundamental challenge in the physical, biological, and data sciences~\cite{newman2018network,peixoto2025network,nitzan2017revealing,timme2007revealing, brunton2016discovering}. 

In particular, whether mapping macroscopic brain networks from neural activity~\cite{bullmore2009complex}, deciphering gene regulatory networks (GRNs) from transcriptomic profiles~\cite{karlebach2008modelling}, or tracking epidemic spread across communities~\cite{pastor2015epidemic}, accurately inferring the underlying causal topology is a critical prerequisite for systems-level understanding. Only with a highly accurate structural map can we reliably predict how a complex system will respond to targeted perturbations---such as localized brain stimulations, node ablations, or pharmacological treatments---enabling the identification of true causal drivers and therapeutic targets.
However, structure and dynamics are deeply intertwined in complex systems~\cite{gross2008adaptive,barzel2013universality,ghavasieh2024diversity,strogatz2024nonlinear,barzon2024unraveling,sole2001complexity,de2026decoding}, and thus inferring the network structure without explicitly modeling the underlying dynamics can lead to inaccurate and unreliable results. 
When the system is governed by nonlinear, stochastic dynamics, and the observations are noisy and limited in time, the problem becomes particularly severe: multiple distinct configurations of the interactions can produce trajectories that are statistically indistinguishable from the data~\cite{villaverde2014reverse}. These difficulties, broadly referred to as structural non-identifiability, call for inference approaches that go beyond point estimation, instead quantifying and propagating uncertainty over the full space of plausible model configurations.
Another important factor is that when dealing with networks, even small changes in the parameters can lead to large changes in the system's behavior, due to the emergence of collective phenomena and critical transitions~\cite{boccaletti2006complex,dorogovtsev2008critical,pastor2015epidemic, strogatz2024nonlinear}. This leads to highly discontinuous likelihood landscapes, where the maxima are isolated and separated by low-probability barriers. For this reason, in most of the cases, standard incremental optimization methods are not good solutions for such problems.
A natural and principled route to this goal is Bayesian inference, in which a posterior distribution over model parameters is obtained by combining a likelihood function for the observed data with prior knowledge about the system~\cite{peixoto2019network, newman2018network,raimondo2021measuring,peel2022statistical,peixoto2025network}. This approach allows us to quantify uncertainty over the inferred interactions, which is crucial for guiding experimental validation and for making reliable predictions about the system's behavior.

Most of the methodologies proposed to tackle this problem rely on statistical associations between observed variables, such as correlation or mutual information, to infer the presence of interactions. Examples include ARACNE and related mutual-information approaches for GRNs~\cite{margolin2006aracne,marbach2012wisdom}, or functional connectivity methods (like partial correlation and network deconvolution) for brain networks~\cite{smith2011network}. While specialized mechanistic frameworks like Dynamic Causal Modeling (DCM) exist for specific neuroimaging modalities~\cite{friston2003dynamic}, many general approaches struggle to separate direct causality from indirect cascades. Furthermore, most of these methods produce single-point estimates and do not provide measures of uncertainty over the inferred interactions, which is crucial for guiding experimental validation and for making reliable predictions about the system's behavior.
Other methods attempt to learn the posterior distribution over network configurations using Bayesian structure learning techniques, often based on graphical models and directed acyclic graphs (DAGs)~\cite{glymour2019review}. These methods can provide uncertainty estimates, but they typically assume that the underlying system is governed by a DAG structure, which is not always valid in real systems, and thus they cannot properly capture cyclic interactions, which are common in biological systems. Additionally, they often rely on assumptions of stationarity and equilibrium, which are not always valid in real systems, and thus they cannot fully exploit the information contained in non-equilibrium trajectories, which are often the only ones available in real experiments.
Given the inherently cyclic nature of neural systems, it is crucial to consider methods that can capture such interactions. Multiple approaches have been proposed to solve the issue in different fields~\cite{huang2020causal,runge2019detecting}. However, these methods have limitations, such as single point estimation, lack of uncertainty quantification, and reliance on assumptions that may not hold in real systems, such as stationarity and equilibrium.
A recent approach~\cite{atanackovic2023dyngfn} addresses some of these limitations by proposing a Bayesian framework for inferring cyclic interactions from time-series data by modeling the system's latent drift function through either linear solvers or non-linear MLPs. While this method represents a significant step forward, it still does not explicitly define the mechanistic SDE model underlying the dynamics, and thus remains of difficult interpretation, comparison with other mechanistic models of gene regulation and with the risk of data overfitting. Additionally, it does not provide a principled way to select the dynamical parameters of the SDE model, such as noise amplitude and relaxation rates.

Recent advancements in Bayesian network reconstruction, pioneered by Peixoto and colleagues~\cite{peixoto2019network, peixoto2025network}, have established a gold standard for overcoming structural non-identifiability. By leveraging non-parametric Minimum Description Length (MDL) priors and Stochastic Block Models (SBMs), this framework rigorously penalizes model complexity without relying on arbitrary hyperparameters. However, while these methods have been highly successful in reconstructing networks from discrete-state dynamics---such as epidemic spreading or Ising spin models---extending these non-parametric priors to systems governed by continuous, highly non-linear Stochastic Differential Equations (SDEs) remains a major theoretical and computational challenge.

In this work, we propose a mechanistic, ensemble-based Bayesian framework for neural network inference from non-equilibrium time-series. We formulate a dynamical likelihood for trajectories governed by non-linear Stochastic Differential Equations (SDEs)---which acts as an Onsager-Machlup quasi-potential~\cite{onsager1953fluctuations} to rigorously weight non-equilibrium paths---and integrate it with the non-parametric Minimum Description Length (MDL) structural prior and Stochastic Block Model (SBM) developed by Peixoto. This joint structural-dynamical optimization is powered by a highly scalable block-coordinate Markov Chain Monte Carlo (MCMC) sampler.

Crucially, our method does not rely on assumptions of stationarity, enabling the full exploitation of transient trajectories which contain richer causal information. We demonstrate that the fundamental challenge of structural non-identifiability can be overcome by globally constraining kinetic parameters, which forces the inference engine to explain heterogeneous biological fluctuations exclusively through the inferred topology rather than absorbing variance via localized parameter overfitting. By navigating the multimodal posterior landscape, our method recovers a rigorous probabilistic ensemble of biologically plausible networks. We show that this ensemble-based approach significantly outperforms state-of-the-art point-estimates, providing robust and reliable predictions for downstream perturbation analysis~\cite{peel2022statistical}.

As a primary application, we start from a nonlinear SDE model with sigmoidal activation functions---capturing saturation, oscillatory, and noise-driven regimes---and perform inference on synthetic networks. We show that the ensemble approach provides a more reliable reconstruction of the true network topology than single-point estimation, particularly for predicting the downstream effects of targeted node ablations. To demonstrate the generality and robustness of the framework against severe generative model mismatch, we additionally apply the inference engine to simulated trajectories from a macroscopic Wilson-Cowan neural rate model, successfully recovering the underlying excitatory and inhibitory neural topology.

\begin{figure*}
    \centering
    \includegraphics[width=\textwidth]{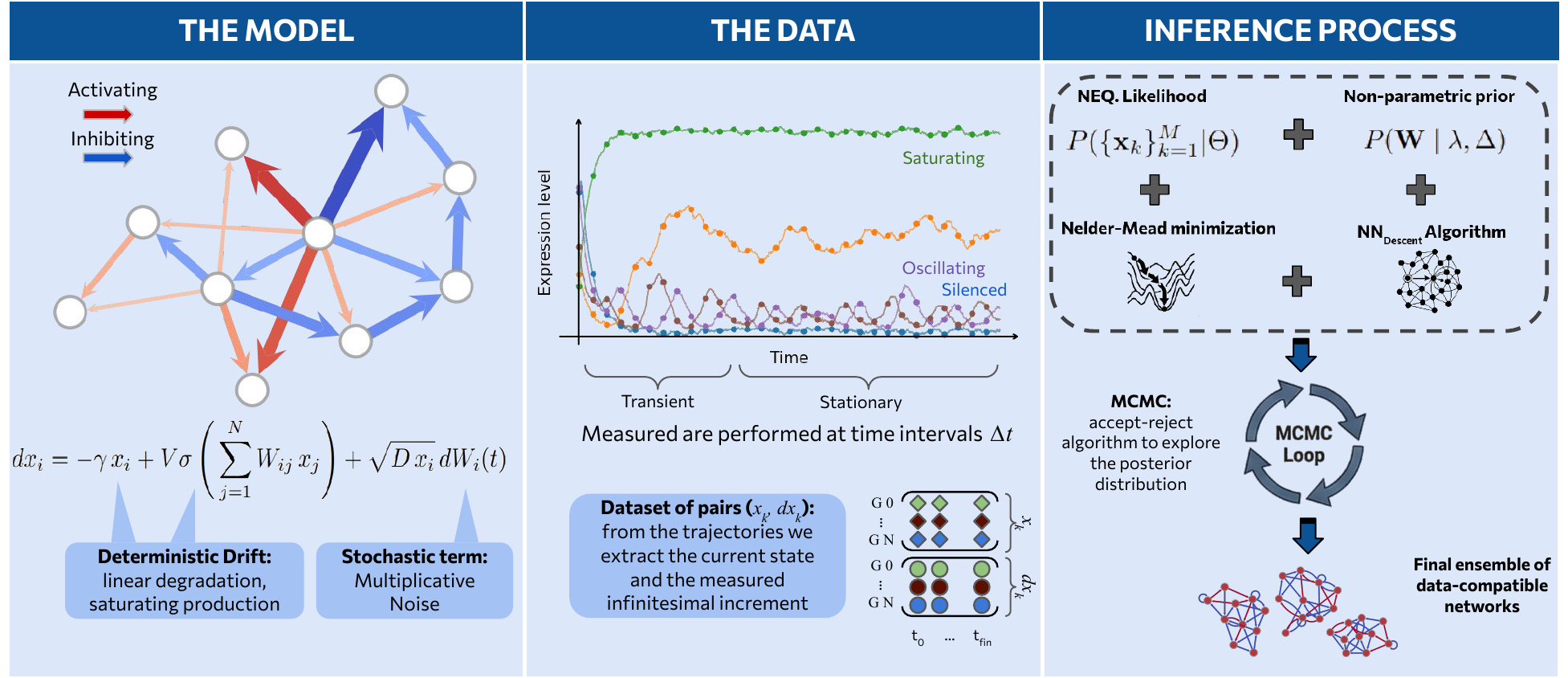}
    \caption{\textbf{Schematic overview of the proposed Bayesian inference framework.} In the left panel, we show the network topology and the model used to generate the synthetic data. In the central panel, we show an example of trajectory generated from the model, and the corresponding dataset of pairs $(\mathbf{x}_k, \mathbf{dx}_k)$ of states and infinitesimal increments $\mathbf{dx}_k = \frac{\Delta \mathbf{x}_k}{\Delta t}$, where $\Delta \mathbf{x}_k = \mathbf{x}_k - \mathbf{x}_{k-1}$, that can be used to perform inference on the parameters of the model. In particular we distinguish between transient and stationary phases in the temporal evolution of the system, and we show the possible different behaviors that can be observed in the trajectories, such as oscillations, saturation and low signal. In the right panel, we show a schematic representation of the inference procedure. First, we perform a node-wise analysis of the likelihood landscape to find an ensemble of configurations for the incoming edges of each node that are compatible with the observed data. Then, we sample from the configurations of the incoming edges of each node to build an ensemble of configurations for the whole network by comparing the emerging global network properties with the prior knowledge about the system. The output of the procedure is an ensemble of configurations for the parameters, each weighted by its posterior probability, that can be analyzed to extract probabilistic measures on the system under analysis.}
    \label{fig:framework}
\end{figure*}

\section{Bayesian Network Inference}


When trying to approach the inverse problem of network reconstruction from some data $\mathbf{X}$, the most rigorous and transparent approach is using Bayesian inference:

\begin{equation}
    P(\mathbf{W},\theta|\mathbf{X}) = \frac{P(\mathbf{X}|\mathbf{W}, \theta)P(\mathbf{W}, \theta)}{P(\mathbf{X})}
\end{equation}

Using this equation, we suppose that the data we have are generated by a specific generative model, described by the likelihood function $P(\mathbf{X}|\mathbf{W}, \theta)$, where $\mathbf{W}$ is the encodes the network structure and weights and $\theta$ are the extra parameters of the generative model. 
In this work we will focus on data organized in time series, so, if we have a network of $N$ nodes, and the data are organized in $M$ time points, we can write the data as a matrix $\mathbf{X} = \{\mathbf{x}_{t_1}, \mathbf{x}_{t_2}, ..., \mathbf{x}_{t_M}\}$ of size $M \times N$, where $\mathbf{x}_m$ is a vector of size $N$ that encodes the state of the nodes at time $t_m$. In this case, the likelihood function can be factorized as:

\begin{equation}
    P(\mathbf{X}|\mathbf{W}, \theta) = \prod_{m=1}^{M-1} P(\mathbf{x}_{t_m+\Delta t}|\mathbf{x}_{t_m}, \mathbf{W}, \theta)
    \label{eq:general_likelihood}
\end{equation}

From a statistical mechanics perspective, the negative log-likelihood of these continuous-time trajectories can be formally interpreted as an action functional, closely related to the Onsager-Machlup quasi-potential~\cite{onsager1953fluctuations}. For non-equilibrium systems that lack a well-defined stationary distribution, this path-integral formulation is theoretically crucial: it provides a rigorous thermodynamic weight for the transient trajectories, allowing the inference engine to map the complex dynamical landscape by evaluating the most probable paths of the system over time.

In most of the realistic scenarios, the data we have are scarce and noisy, and the generative models are complex and high-dimensional. For this reasons, when we try to perform inference on the generative model, we can encounter two main problems: non-identifiability and overfitting. Non-identifiability occurs when multiple distinct configurations of the interactions can produce trajectories that are statistically indistinguishable from the data. Overfitting occurs when the model is too complex and fits the noise in the data rather than the underlying signal. 

To overcome these issues we need to incorporate an extra term which acts as a regularizer, penalizing complex models and incorporating prior knowledge about the system. This term is the prior distribution $P(\mathbf{W}, \theta)$, that enables the inference process to be guided towards more plausible configurations, and to avoid overfitting.


One way to interpret the Bayesian inference framework is through the lens of Minimum Description Length (MDL) principle~\cite{rissanen1978modeling}. Taking the logarithm of Bayes' theorem, we can rewrite it as:

\begin{equation}
    \log P(\mathbf{W},\theta|\mathbf{X}) = \log P(\mathbf{X}|\mathbf{W}, \theta) + \log P(\mathbf{W}, \theta) - \log P(\mathbf{X})
\end{equation}

In this way the inference problem can be seen as a maximum compression problem, where we want to find the network configuration that requires the least amount of information to describe the data. This is a crucial aspect of Bayesian inference, as it clearly states the goal of the inference process: to find the network configuration that best explains the data while being as simple as possible.


Once we have defined both the generative model and the prior distribution, we can find the configuration that is able to best compress the data, which is called the Maximum A Posteriori (MAP) estimate. However, in many cases, the MAP estimate is not sufficient to capture the full complexity of the inference problem. In particular, the best network configuration often does not correspond exactly to the true underlying network, due to data scarcity, noise and other aspects that cannot be completely comptured by the generative model. 

This aspect is particularly important in contexts where the mis-specification of a single edge can have a relevant impact on subsequent system analysis. This is particularly relevant in complex networked systems (such as biological pathways, ecological webs, or neural circuits), where the inferred network is frequently used to perform downstream perturbation analysis, and the mis-specification of a single causal edge can lead to completely wrong macroscopic predictions. 

For this reason, it is crucial to consider the full posterior distribution over network configurations, rather than just the MAP estimate. This allows us to capture the uncertainty in the inference process and to generate a probabilistic ensemble of plausible network configurations, which can be used to perform more robust and reliable analyses. This can be acheived by sampling from the posterior distribution using Markov Chain Monte Carlo (MCMC) methods, which allow us to generate a representative set of network configurations that are consistent with the generative model and the model's constraints.

\subsection{Capturing Saturation and Relaxation in Networked Systems}


As a generative model for the data, we consider a system of stochastic differential equations (SDEs) that describe the dynamics of a network of interacting nodes. This is the most general and flexible framework for modeling complex systems, as it allows us to capture both the deterministic dynamics of the system and the intrinsic stochasticity that characterizes the aquired data. 
As a representative case study, we focus on the following model:

\begin{equation}
    dx_i = -\gamma \, x_i + V \sigma \, \Bigg(\sum_{j=1}^{N} W_{ij}\, x_j \Bigg) +  \sqrt{D \, x_i} \, dW_i(t)
\end{equation}

Although simple, this model is able to capture the universal features of bounded complex networks, namely, intrinsic exponential relaxation and bounded non-linear coupling. Such features are common to many real-world systems, including ecological systems, macroeconomic networks, gene regulation, and brain networks. While these features map universally to population dynamics (e.g., carrying capacity in ecology or mRNA degradation in cells), we will frequently use macroscopic neural population dynamics as an illustrative framework throughout this work. In this context, the exponential decay term represents intrinsic membrane relaxation, the saturating non-linear activation represents the limited capacity of firing rates, and the multiplicative noise captures the intrinsic stochasticity characterizing natural systems.

In these kinds of complex systems, the correct inference of the interaction matrix $W$ is crucial for understanding the underlying regulatory mechanisms and for making predictions about the system's behavior under different environmental conditions or perturbations. For this reason, the Bayesian inference framework is particularly well-suited for this task, as it allows to produce a probabilistic ensemble of plausible network configurations that can be used to perform more robust and reliable analyses.


A particular aspect of the model that we consider is the choice of the multiplicative noise term, which is modeled as a Cox-Ingersoll-Ross (CIR) process~\cite{cox1985theory}. This choice is motivated by the fact that many complex systems in nature exhibit bursty, avalanche-like dynamics where fluctuations are inherently proportional to the current state of the system---such as demographic noise in ecology, transcriptional bursting in cells, or avalanches in neural activity. This state-dependent burstiness is naturally captured by the multiplicative CIR noise term. In contrast, additive noise would fail to capture this fundamental physical property, as it introduces fluctuations that are entirely independent of the system's macroscopic state.



\subsection{Theoretical Validation: Exact Transition Moments}


While computationally trivial, the Euler-Maruyama linear approximation~\cite{kloeden1992numerical} is mathematically guaranteed to become unstable at large measurement intervals ($\Delta t$), which are common in real-world transcriptomic time-series (e.g., hours between sampling). Under large $\Delta t$, the linear tangent completely ignores the exponential relaxation inherent to the $-\gamma x_t$ degradation term. If the degradation rate is fast ($\gamma \Delta t > 1$), this results in severe mathematical overshoot, where the predicted mean may oscillate and drop below zero. Furthermore, EM assumes the variance grows linearly without bound ($2D x_t \Delta t$), ignoring the variance saturation caused by mean-reverting dynamics.


To construct a theoretical stress-test and validate the robustness of this discretization, we introduce an exact analytical integration approach predicated on a zero-order hold assumption. We assume that the aggregate regulatory input from the rest of the network, $m_i(t)$, remains piecewise constant over the short interval $[t, t+\Delta t]$. Under this assumption, the non-linear coupling term $V \, \sigma(m_i(t))$ simplifies to a constant scalar $b$.

The governing SDE for a single isolated node during this time step reduces to a strictly linear stochastic differential equation:

\[ dx_t = (-\gamma x_t + b) dt + \sqrt{2D x_t} dW_t \]

By isolating the node's state variable $x_t$, the SDE is mathematically equivalent to a continuous CIR process, which permits an exact analytical solution for its transition moments regardless of the magnitude of $\Delta t$. 

    
By defining the deterministic steady-state asymptote as $\mu_\infty = \frac{V \, \sigma(m_i(t))}{\gamma}$, the exact expected mean is strictly bounded by the exponential relaxation:

\[ \mu(t+\Delta t) = \mu_\infty + \big(x_i(t) - \mu_\infty\big)e^{-\gamma_i \Delta t} \]

And the exact integrated variance accumulated over the interval $\Delta t$ correctly saturates to a finite limit:

\begin{equation}
    \begin{aligned}
    \Sigma(t+\Delta t) &= \frac{2D_i}{\gamma_i}\big(x_i(t) - \mu_\infty\big)\Big(e^{-\gamma_i \Delta t} - e^{-2\gamma_i \Delta t}\Big) \\
    &\quad + \frac{D_i \mu_\infty}{\gamma_i}\Big(1 - e^{-2\gamma_i \Delta t}\Big)
    \end{aligned}
\end{equation}

The full analytical derivation of these moments via Itô's Isometry is provided in the Supplementary Material.

\subsection{Inference engine}


Once the generative model is defined, we employ an extension of the Markov Chain Monte Carlo (MCMC) inference engine developed by Peixoto~\cite{peixoto2025network}, implemented via the \textit{graph-tool} library. 
This method incorporates a hierarchical, non-parametric Minimum Description Length (MDL) prior, which goal is to rigorously penalize model complexity without relying on weight shrinkage. It achieves this via weight quantization: rather than treating every non-zero edge as having an infinitely precise, unique weight (which would require infinite information to describe), the algorithm is forced to compress the network's continuous weights into a highly redundant, finite vocabulary of discrete categories. 

This adaptive compression is formalized by marginalizing the prior over several latent structural and parametric variables:
\begin{equation}
    \begin{aligned}
        P(\mathbf{W} \mid \lambda, \Delta) &= \sum_{z,K,\mathbf{A},E} P(\mathbf{W} \mid z, \mathbf{A})\, P(z \mid \lambda, \Delta, K)\, \\
        &\quad\times P(K \mid \mathbf{A})\,P(\mathbf{A} \mid E)\,P(E)
    \end{aligned}
    \label{eq:mdl_prior}
\end{equation}

where $\mathbf{W}$ represents the continuous interaction weights of the network, and $\mathbf{A}$ is its corresponding binary adjacency matrix containing exactly $E$ non-zero edges. The parameter $K$ denotes the total number of unique, discrete weight categories into which the edges are grouped, while $z = \{z_1, \dots, z_K\}$ represents the set of continuous values for those $K$ categories.

This hierarchical formulation allows the inference engine to dynamically adjust the complexity of the model. Several of the marginal distributions represent simple uniform priors that serve as baseline complexity penalties: $P(E)$ is uniformly distributed over the possible number of edges and $P(K \mid \mathbf{A})$ is uniform over the number of discrete weight categories ($K \le E$).

The term responsible for the sparse structure of the network is:

\begin{equation} 
    P(\mathbf{A} \mid E) = \frac{\delta_{\sum_{i,j}A_{ij}, E}}{\binom{N^2}{E}} 
\end{equation}

which penalizes topologies with large number of edges, leading to small description length for sparse graphs.

The other two terms govern how the continuous weights are compressed into the discrete vocabulary $z$. The first one, $P(\mathbf{W} \mid z, \mathbf{A})$, quantifies how much the number of paramters categories impact on the compression level of the data.
If $m_k$ represents the number of edges assigned to a specific weight category $k$, this probability distribution yields a multinomial combinatorial term proportional to $\frac{\prod_k m_k!}{E!}$. From an MDL perspective, this term heavily rewards parameter reuse. Because the inverse of the multinomial coefficient maximizes when many edges share the exact same weight category (e.g., $m_k$ is large), the algorithm is strongly incentivized to minimize $K$ and group the weights into a small, highly redundant vocabulary.

The second term, $P(z \mid \lambda, \Delta, K)$, prevents the absolute values of the weight categories from diverging. They are drawn from a quantized Laplace distribution defined by the scale parameter $\lambda$ and a numerical precision $\Delta$, yielding a probability proportional to $e^{-\lambda \sum_{k=1}^K |z_k|}$. This does not yeald to strong shrinkage of the weights, making it a more robust alternative to the common LASSO regularization.

Alongside to these terms, the prior could also include the Stochastic Block Model (SBM) with degree correction, which allows the algorithm to exploit the modular structure of the network to further compress the description length. This is particularly useful in the context of neural networks, where regions are often organized into functional modules that share similar regulatory patterns, and for the case scenario considered in this work, is able to classify between the master regulators and the downstream targets, improving the overall inference performance.

All these elements compose a non-parametric prior that is able to adaptively penalize model complexity, while still allowing for a rich and flexible representation of the network structure and weights. This is crucial for avoiding overfitting and ensuring that the inferred networks are both biologically plausible and statistically robust.


To optimize the joint posterior of the network topology and the global kinetic parameters, we employ a block-coordinate optimization scheme. The algorithm alternates between two phases: first, the discrete topological space (edges and block assignments) is sampled stochastically using Metropolis-Hastings MCMC sweeps. The sampling scheme for the topology is based on the $NN_{Descent}$ algorithm~\cite{dong2011efficient}, which is able to solve the combinatorial optimization problem in a subquadratic time complexity of order $O(N\,log^2 N)$, instead of the naive $O(N^2)$ time complexity for standard gradient descent algorithms. On the other hand, the classes of parameters values are obtained by a system of merging and splitting of the weight classes and random bisection to obtain new values for the weight categories.

Second, the global continuous kinetic parameters $\theta = \{\gamma, V, D\}$ are deterministically optimized conditional on the current topology using the Nelder-Mead simplex algorithm~\cite{nelder1965simplex} to minimize the total description length. This choice was made because the likelihood function is convex with respect to the kinetic parameters, and given a fixed topology and weight categories, the optimization problem is well-posed and can be solved efficiently using a deterministic algorithm.

This hybrid approach allows the engine to efficiently navigate the highly combinatorial structural space while quickly relaxing the continuous dynamics into their local optimum.

\begin{figure*}
    \centering
    \includegraphics[width=\textwidth]{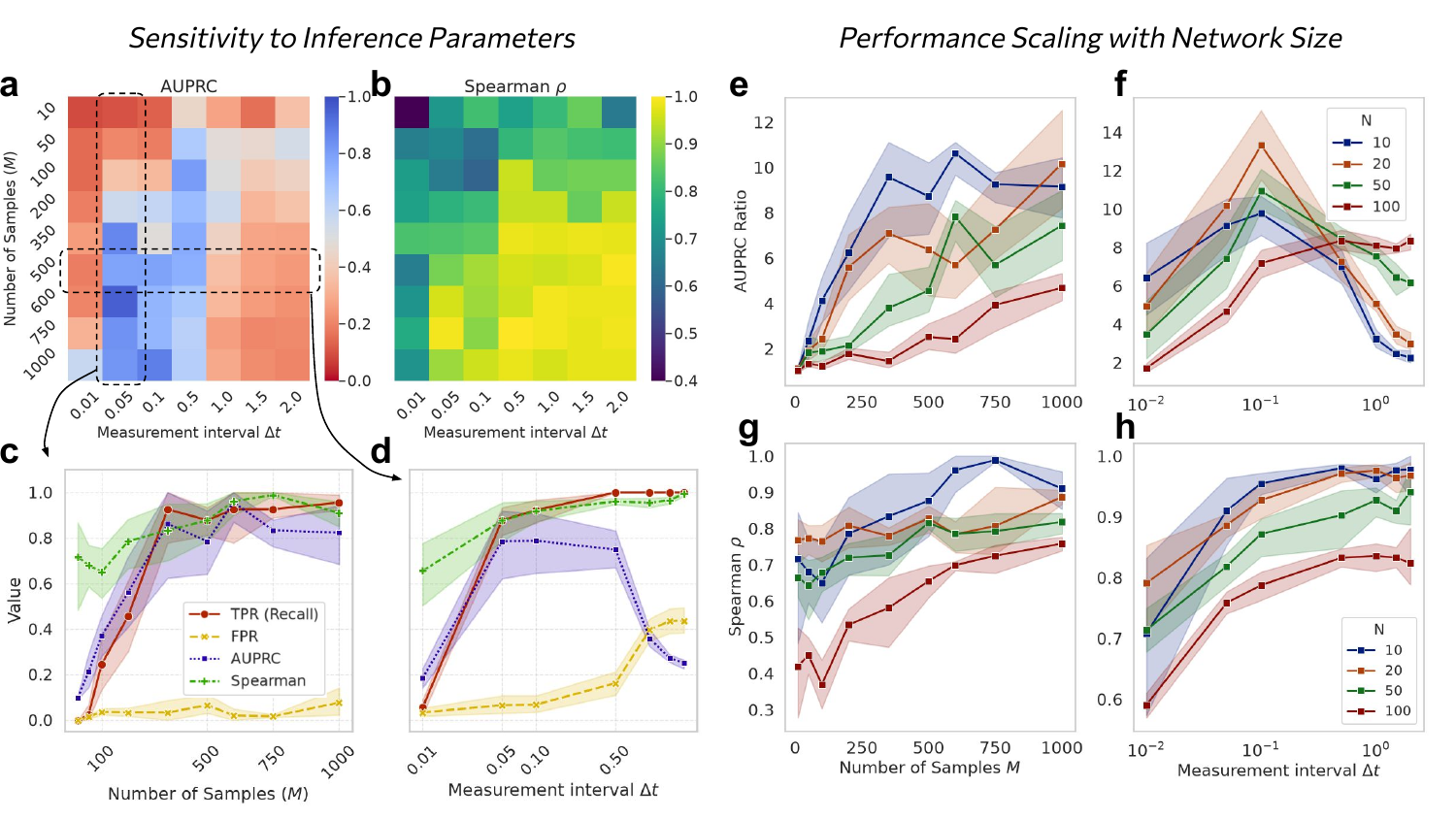}
    \caption{\textbf{Inference performance analysis:} In panel a and b we present the grid search results to evaluate the inference performance as a function of the number of sampled pairs $M_{pairs}$ and the sampling interval $\Delta t$. a reports the AUPRC, which quantifies the topological accuracy of the probabilistic inferred network, while b reports the Spearman correlation coeffient, which quantifies the accuracy of the inferred regulatory weights. The results are for network of size $N=10$ and for the transient regime. In panel c and d we report respectively the slices of the grid search results both trough the variation of sampling interval $\Delta t$ (c) and the number of sampled pairs $M_{pairs}$ (d). In particular, in panel c we can see that the inference performance reaches a maximum for intermediate values of $\Delta t$, and after that point degrading them as a result of an increment of the FPR, while in panel d we can see that in general the inference performance improves with the number of sampled pairs $M_{pairs}$ until reaching a saturation point. 
    We can see that, while for large measurement intervals $\Delta t$ the topological inference performance is in general poor, the inference of the regulatory weights remains robust, probably due to a overfitting effect that tends to overestimate the number of active edges, leading to a high FPR but still capturing the correct regulatory weights.
    In the panels e, f, g and h we report the scaling of the inference performance with the network size $N$ for the transient regime. In particular, in panel e, f we consider the AUPRC Ratio, that quantifies how much better the inference performance is with respect to a random baseline, while in panel g, h we consider the Spearman correlation coefficient, that quantifies the accuracy of the inferred regulatory weights. In particular, in panel e and g we report the results for a fixed sampling interval $\Delta t=0.05$, while in panel f and h we report the results for a fixed number of sampled pairs $M_{pairs}=500$. What we can see is that the inference performance is able to produce good results even for large networks, given a sufficient number of sampled pairs and a suitable sampling interval. In particular, from the panels f, h, we can see that larger networks are more robust to the choice of the sampling interval $\Delta t$, reaching a plateau of performance for a wider range of values. While the topological reconstruction is in general more difficult for larger networks, from plots g, h we can see that the inference of the regulatory weights is in general more robust to the network size, producing comparable results across different network sizes.}
    \label{fig:inference_performance}
\end{figure*}

\section{Results}

To systematically evaluate the performance of our Bayesian inference framework, we conducted a comprehensive grid search analysis on synthetic, scale-free complex networks modeled via non-linear SDEs. The experimental setup was designed to explore a wide range of realistic observational conditions by varying the network size ($N$), the number of sampled time-series trajectories ($M_{pairs}$), the sampling interval ($\Delta t$), and the dynamical regime from which the data were collected (transient versus stationary). Unless otherwise specified, the networks were generated as directed scale-free graphs using the Price model. We chose scale-free topologies as our primary benchmark for two reasons: first, they represent the canonical standard for macroscopic complex systems across biology; second, their highly-connected "master regulators" (hubs) drive massive, system-wide indirect correlation cascades. Successfully resolving these networks proves the Bayesian engine can overcome severe topological confounding, distinguishing true sparse causal drivers from dense indirect effects. Trajectories were simulated under moderate intrinsic noise. The results of this extensive analysis demonstrate the robustness of the ensemble-based approach in recovering the underlying true network topology. (Full details regarding the generative model parameters and integration schemes are provided in the Supplementary Materials).

The structural reconstruction of the causal topology demonstrates a strong dependence on the experimental inference settings, particularly the temporal measurement resolution ($\Delta t$) and the number of available sample pairs ($M$).

While increasing the temporal measurment interval ($\Delta t$) initially improves structural recovery, given that for small $\Delta t$ the finite-difference derivative estimates are dominated by noise, we observe a sharp decline in performance for excessively large $\Delta t$. As we can see in Fig.~\ref{fig:inference_performance}c, the True Positive Rate (TPR) remains high, but the False Positive Rate (FPR) begins to increase, leading to a degradation of the overall topological inference performance, represented by the Area Under the Precision-Recall Curve (AUPRC). This is primarily driven by three factors: (1) the accumulation of stochastic noise which decorrelates consecutive states and obscures the deterministic causal signal; (2) the propagation of signals through intermediate nodes, which causes indirect regulatory cascades to be falsely identified as direct interactions; and (3) the breakdown of the discrete-time finite-difference approximation of the underlying continuous-time stochastic differential equations.

The relationship between the sample size ($M$) and performance is heavily modulated by the measurement interval ($dt_{meas}$). When $dt_{meas}$ is sufficiently small, the discrete-time finite-difference model accurately approximates the true continuous dynamics. In this regime, providing more data ($M$) simply increases statistical confidence, causing the inference performance to asymptotically saturate at a high level. Conversely, when $dt_{meas}$ is large, the discrete-time approximation becomes structurally flawed. At low $M$, the algorithm lacks the statistical power to notice this model mismatch and still captures the dominant true edges. However, as $M$ grows very large, the algorithm is forced to explain away the systemic discretization errors. With immense statistical confidence, it introduces spurious interactions to fit the artifactual dynamics, causing the False Positive Rate to rise and the overall performance to degrade. Thus, massive datasets ($M \to \infty$) paradoxically harm performance if the temporal resolution ($dt_{meas}$) is fundamentally misaligned with the system's dynamics.

In particular, the grid search revealed that the method performs best when operating on non-equilibrium transient trajectories and when employing the exact analytical transition probability, which successfully mitigates the severe discretization errors that typically plague naive approximations at large sampling intervals. In the following subsections, we delve into the specific insights drawn from these experiments.

Finally, we evaluated the scalability of the inference framework with respect to the network size ($N$). As the number of nodes increases, the dimensionality of the causal inference problem expands quadratically ($O(N^2)$), making structural reconstruction strictly more challenging. Consequently, the absolute Area Under the Precision-Recall Curve (AUPRC) naturally drops for larger networks, driven primarily by the decrease in baseline density. However, when evaluating the AUPRC Ratio, the model maintains performance vastly superior to random chance across all tested sizes. Interestingly, the inference of quantitative regulatory weights (measured via Spearman correlation) is remarkably robust to network size, yielding comparable accuracy regardless of $N$. Furthermore, empirical results indicate that larger networks are actually more robust to the choice of the measurement interval ($\Delta t$), maintaining a high plateau of inference performance across a significantly wider range of temporal resolutions than smaller networks. This confirms that the Bayesian framework robustly scales to larger, highly connected complex systems without a catastrophic loss of predictive power. To ensure these scaling properties were not strictly dependent on the presence of hubs, we replicated the entire grid search on homogenously random Erdős-Rényi (ER) networks (see Supplementary Materials). A comparative analysis revealed a fundamental structural trade-off: while the lack of dominant hubs in ER networks actually improves the quantitative inference of regulatory weights at small scales ($N \le 20$), Scale-Free topologies proved significantly more robust to dimensional scaling. The hubs in Scale-Free graphs act as macroscopic dynamical anchors, allowing the algorithm to maintain a higher relative topological accuracy (AUPRC Ratio) in large, high-dimensional spaces where the completely homogeneous noise of ER networks becomes overwhelmingly difficult to parse.

\subsection{The necessity of simple models: the problem of overfitting}

In the governing SDE model (Eq.~1), the dynamical parameters $\theta = \{\gamma, V, D\}$ are assumed to be global and shared across all nodes in the network. To evaluate whether relaxing this assumption improves reconstruction, we compared this global parameterization against a ``local'' mode, in which each node $i$ is assigned independent parameters $\theta_i = \{\gamma_i, V_i, D_i\}$.

As shown in Fig.~\ref{fig:benchmarking}a, allowing parameters to vary locally leads to a clear drop in performance, characterized by lower AUPRC ($0.5 \pm 0.2$ vs $0.8 \pm 0.1$), higher Speraman correlation ($0.38 \pm 0.12$ vs $0.48 \pm 0.03$), and an increased rate of false positives. This behavior arises from an identifiability problem between intrinsic node kinetics and network coupling: variations in a node's trajectory can be fitted either by adjusting its local kinetic rates ($\gamma_i, V_i$) or by introducing spurious regulatory edges ($W_{ij}$). In the local mode, increasing the sample size or reducing observational noise does not resolve this issue; instead, the likelihood fits residual trajectory fluctuations by adding false edges.

Constraining the kinetic parameters to be global resolves this degeneracy. By establishing a uniform kinetic baseline across all nodes, the model is forced to attribute differences in node trajectories to the interaction topology $\mathbf{W}$. Consequently, this global constraint serves as an effective regularizer, making the framework well-suited for systems where the collective dynamics are predominantly shaped by network connectivity rather than individual kinetic heterogeneity.

\subsection{Benchmarking against state-of-the-art inference paradigms}

This kind of model is able to help avoid overfitting also when compared to other state-of-the-art inference models. We performed a comprehensive benchmarking study where our proposed Bayesian stochastic differential equation (SDE) approach was compared against an extensive suite of nine distinct causal discovery and dynamical system paradigms: dynGENIE3~\cite{huynh2018dyngenie3} (non-parametric tree-based inference), Sparse ODE via Lasso~\cite{tibshirani1996regression, bonneau2006inferelator} and Elastic Net VAR~\cite{zou2005regularization} (sparse parametric linear recovery), Granger Causality~\cite{granger1969investigating, shojaie2010discovering} (classical statistical temporal prediction), PCMCI~\cite{runge2019inferring} (information-theoretic causal discovery), Dynamic Bayesian Networks~\cite{husmeier2003sensitivity} (discrete-time probabilistic structure learning), Gaussian Process ODEs~\cite{calderhead2008accelerating} (non-parametric Bayesian regression), PySINDy~\cite{kaptanoglu2021pysindy} (modern data-driven sequential thresholding), and Neural ODEs~\cite{chen2018neural} (highly parameterized continuous-time deep learning). Over independent replicates of simulated scale-free networks, we assessed the models on their structural recovery and predictive generalization. 

The results reported in Fig.~\ref{fig:benchmarking}d indicated a strong universal tendency for standard baselines to overfit the highly transient training data whenever given excessive parameter flexibility. For instance, while dynGENIE3 and Neural ODEs could frequently fit the training trajectories perfectly, they exhibited a massive discrepancy between training and out-of-sample test prediction errors, indicating severe overfitting to noise, and failed to recover the structural topology. Information-theoretic and statistical methods like PCMCI and Granger Causality also struggled significantly, as they typically require long, stationary time-series rather than short transient snapshots. Conversely, our Graph-Tool SDE implementation provided robust structural inference (AUPRC = $0.83\pm0.05$) while maintaining almost identical training and test errors ($\Delta \approx 0.004\pm0.004$). This parity indicates strong trajectory compression and out-of-sample generalization, demonstrating that the central advantage of the method lies in the Bayesian framework's inherent capacity to prevent overfitting and accurately recover the underlying causal interactions.

\subsection{Bayesian Topology Inference is Robust to SDE Discretization Errors}

\begin{figure*}
    \centering
    \includegraphics[width=\textwidth]{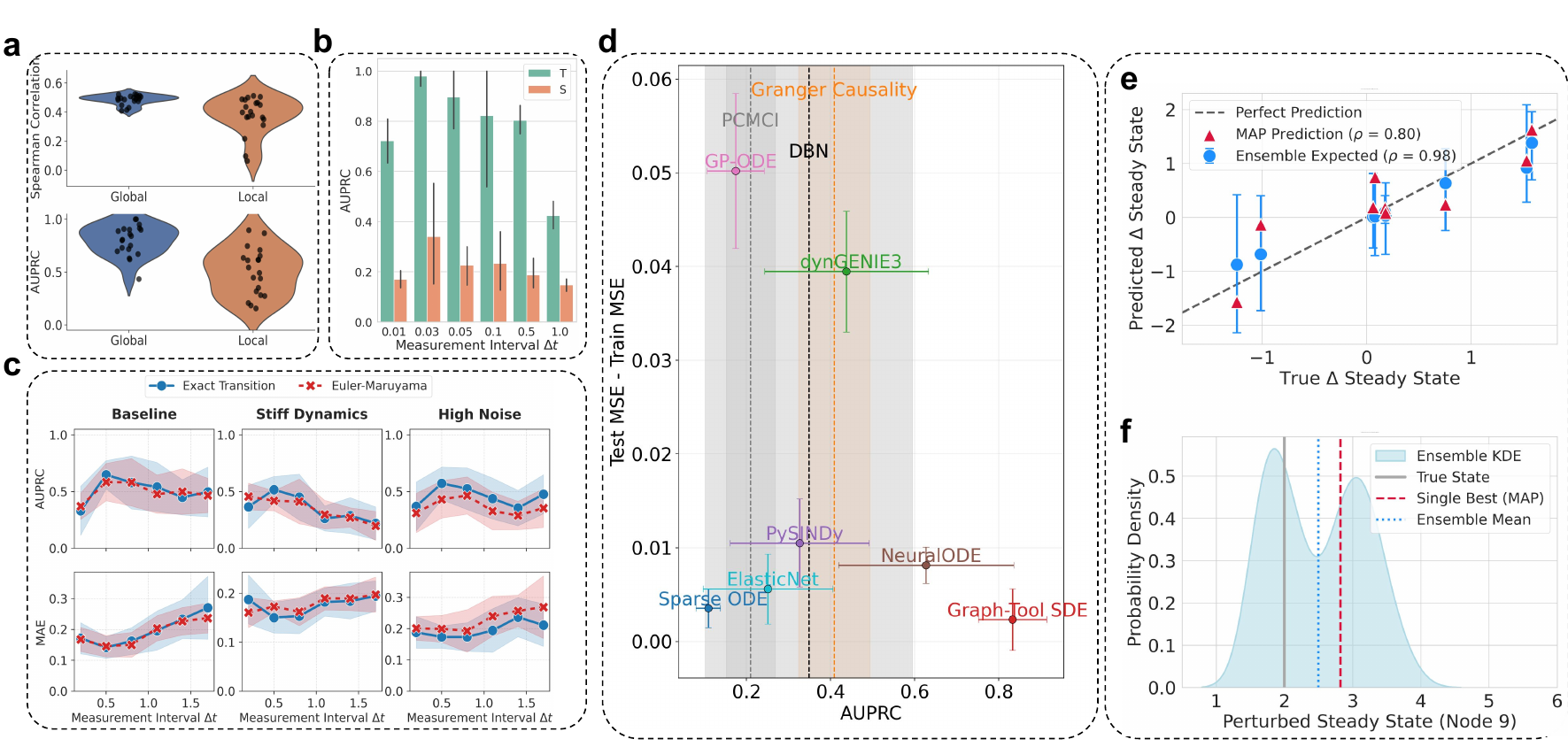}
    \caption{\textbf{Benchmarking against canonical inference paradigms.} 
    In panels a, b, and c, we report the robustness of the inference engine across experimental regimes: 
    (a) parameter regularization, comparing node-specific (local) vs. shared (global) kinetic parameters $\theta = \{\gamma, V, D\}$; 
    (b) non-equilibrium dynamics, comparing transient vs. stationary trajectories; and 
    (c) discretization sensitivity, comparing exact transition moments against the Euler-Maruyama (E-M) approximation. 
    In panel d, we benchmark the proposed Bayesian SDE approach against an extensive suite of nine canonical inference paradigms spanning tree-based, sparse regression, information-theoretic, deep learning (Neural ODEs), and dynamical systems models (see Materials and methods and Supplementary Material for full details). Performance is evaluated by topological accuracy (AUPRC) and generalization gap (train vs. test MSE). While overparameterized baselines exhibit severe overfitting or reduced topological recovery, the Bayesian SDE achieves high structural accuracy with minimal generalization gap. 
    In panels e and f, we assess downstream perturbation prediction following in silico node ablation: 
    (e) correlation between true and predicted changes in downstream target steady states comparing the MAP point-estimate ($\rho = 0.800$) against the posterior ensemble ($\rho = 0.983$); and 
    (f) Kernel Density Estimation (KDE) log-likelihood of the systemic perturbed state under MAP vs. ensemble predictive distributions.}
    \label{fig:benchmarking}
\end{figure*}

The Euler-Maruyama (E-M) approximation~\cite{kloeden1992numerical} is a widely used method for discretizing stochastic differential equations (SDEs) due to its simplicity and computational efficiency. However, it is well-known that E-M can introduce significant discretization errors, particularly when the sampling interval $\Delta t$ is large relative to the system's intrinsic relaxation timescales.
To assess the impact of these discretization errors on network inference, we conducted a systematic benchmarking study comparing the performance of our inference engine using both the exact analytical transition moments and the E-M approximation across a range of sampling intervals.
By looking at the results in Fig.~\ref{fig:benchmarking}c, we found that both methods lead to comparable inference performances, even at large $\Delta t$ where E-M is theoretically expected to fail. In particular, we tested the inference engine on both the standard conditions used for the grid search and on stress-test conditions, in one case significantly increasing the degradation rate $\gamma$ forcing a stiff dynamics, and in the other case significantly increasing the noise level $D$ leading to a highly stochastic dynamics. In both cases, the inference engine produced comparable results using either the exact analytical transition moments or the E-M approximation, demonstrating that the inference engine is robust to the discretization errors introduced by E-M.

A possible explanation for this result is that the inference engine is able to compensate for the discretization errors introduced by E-M by adjusting the global kinetic parameters $\theta = \{\gamma, V, D\}$. In particular, the optimizer can tune these parameters to effectively absorb the discrepancies between the E-M approximation and the true dynamics, allowing the inference engine to still recover a plausible network topology.

This suggests that, for the purpose of network inference, the E-M approximation is sufficiently accurate and can be used as a computationally cheaper alternative to the exact analytical method.

\subsection{Working outside stationarity conveys more information}


Our Bayesian SDE approach is particularly well-suited for analyzing non-equilibrium transient trajectories, which are often the only type of data available in real-world biological experiments. In contrast, many existing methods rely on stationary or equilibrium assumptions, which can lead to misleading inferences when applied to transient data. By evaluating the Onsager-Machlup quasi-potential of the system's active relaxation, our framework captures the directed causal forces that are otherwise indistinguishable from symmetric noise at equilibrium.

To illustrate this, we analyze the performance of our method on synthetic datasets generated from both transient and stationary regimes. The results reported in Fig.~\ref{fig:benchmarking}b show that the inference engine achieves significantly higher accuracy when operating on transient trajectories, as these contain richer causal information about the underlying network structure. This holds for all the network sizes and sampling intervals tested, demonstrating the robustness of the approach across a wide range of conditions.
In contrast, stationary data often lead to systematically worse results, probably due to the strong presence of ambiguous correlations that can confound the inference process.

This can be particularly useful in the context of many real-world systems, where the system is often perturbed by external stimuli or environmental changes, leading to transient responses that reveal the causal relationships between system's components.

\subsection{Toward an ensemble perspective}

\paragraph{Probabilistic evaluation of inference results\\}

Standard network inference methods typically yield a single, optimal point-estimate of the network topology (MAP-estimate). However, complex systems are heavily under-determined, meaning many distinct network structures can explain the observed data equally well. To address this, we approach the inference problem probabilistically using Markov Chain Monte Carlo (MCMC) sampling. Rather than forcing binary present/absent edge decisions, the MCMC produces a posterior distribution of viable network topologies. From this ensemble, we can compute marginal edge probabilities and expected interaction weights, effectively shifting the paradigm from finding the "one true network" to capturing and quantifying the full structural uncertainty inherent in the data.

Evaluating this probabilistic output requires a departure from traditional binary classification metrics. To assess topological accuracy, standard metrics such as the Area Under the Precision-Recall Curve (AUPRC) were adapted by utilizing the marginal edge probabilities as natural confidence scores. However, evaluating the inferred coupling weights presents a deeper challenge; due to parameter non-identifiability in highly non-linear dynamical systems, absolute parameter values often diverge even when the macroscopic dynamic behavior is perfectly captured. Therefore, we evaluated the quantitative reconstruction by focusing on the relative hierarchy of the interaction strengths. We primarily employed Spearman Rank correlation to compare the expected inferred weights against the true network, emphasizing that predicting the correct hierarchy of dominant systemic drivers is dynamically more relevant than exact parameter matching. (A comprehensive formulation of these metrics, alongside distributional evaluations using the Wasserstein distance, is detailed in the Supplementary Materials).

\paragraph{Impact on perturbation analysis\\}


The ultimate biological payoff of this probabilistic framework lies in forecasting the downstream effects of targeted interventions, such as drug treatments or node ablations. Predicting perturbation outcomes is highly sensitive to individual false-positive or false-negative edges; a single incorrect regulatory link in a point-estimate model can catastrophically derail the simulated dynamics. To assess the predictive robustness of our approach, we performed an \textit{in silico} perturbation experiment by simulating a node ablation and evaluating the resulting systemic shift in steady states. We contrasted the predictive power of the full posterior ensemble against the Maximum A Posteriori (MAP) estimate---the single sampled network with the minimum description length.

As shown in Fig.~\ref{fig:benchmarking}e, while the MAP point-estimate generalized poorly to the perturbation due to overfitting the wild-type data ($\rho = 0.800$), marginalizing over topological uncertainty allowed the expected shift of the ensemble to accurately predict the systemic response ($\rho = 0.983$). 

Crucially, propagating an ensemble of networks through non-linear dynamics often yields complex, non-trivial probability distributions for the downstream steady states, making simple point-to-point comparisons insufficient. To properly evaluate these predictions, we employed a Kernel Density Estimation (KDE) log-likelihood metric to measure how well the true perturbed states were supported by the ensemble's predicted probability densities. The necessity of this probabilistic approach was most evident in edge cases such the one depicted in Fig.~\ref{fig:benchmarking}f. 

For this downstream target, the MAP point-estimate erroneously predicted almost no regulatory change ($\Delta \approx -0.13$). However, the true biological response was a massive down-regulation ($\Delta \approx -1.01$). Evaluation of the ensemble's KDE revealed a highly informative, non-trivial probability distribution that explicitly captured this massive down-regulation, placing the true state in a region of high predictive likelihood. This demonstrates that point-estimates can fail catastrophically in downstream predictions, and that resolving and propagating structural uncertainty is critical for reliable perturbation forecasting in biological networks.

\subsection{Generality of the framework: robustness to non-linear model mismatch}

\begin{figure*}
    \centering
    \includegraphics[width=\textwidth]{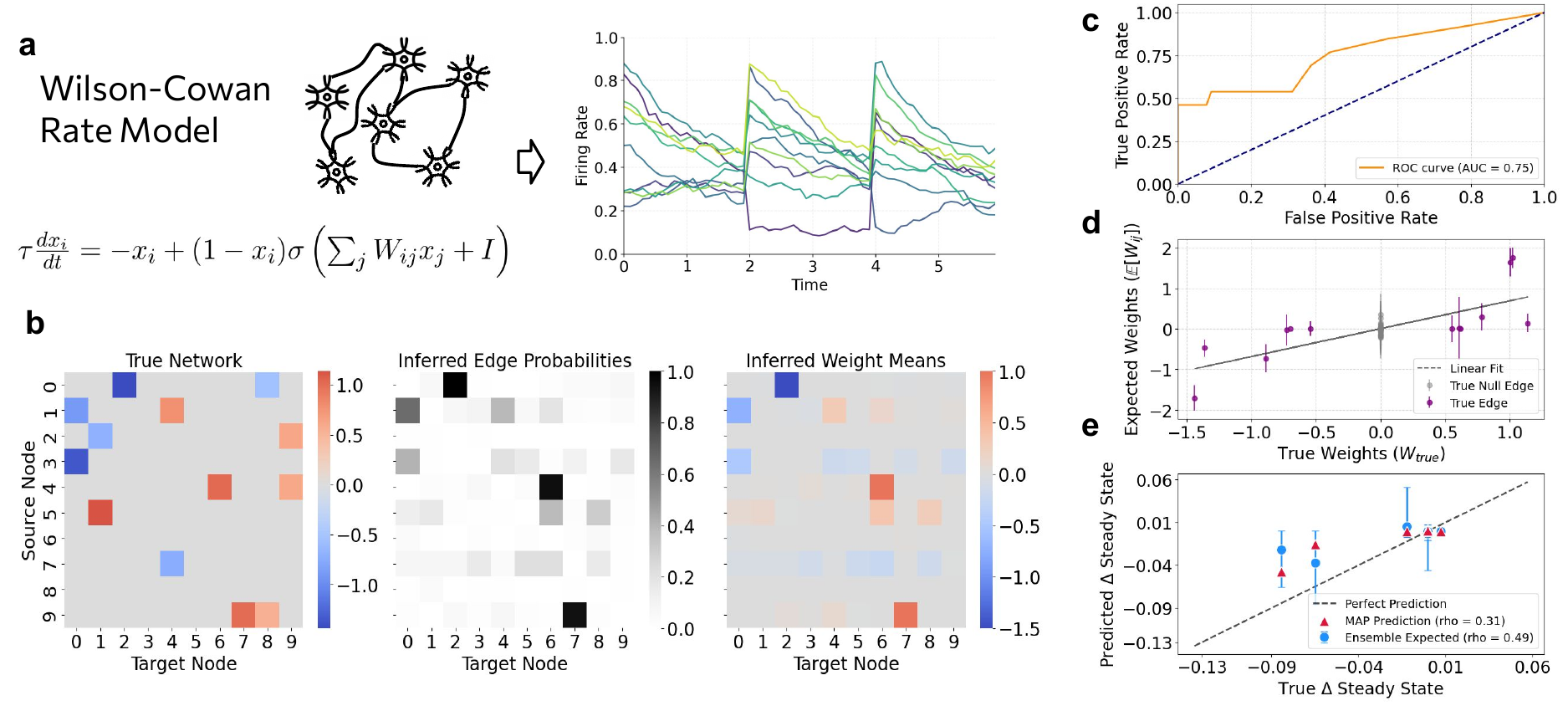}
    \caption{\textbf{Results of Wilson-Cowan Rate Model.} In panel a is presented the schematic of the Wilson-Cowan model, which is a non-linear dynamical system that describes the interactions between excitatory and inhibitory populations of neurons, regulated by a set of non-linear differential equations different from the SDE model used for inference. Because our grid search established that the algorithm excels at quantitatively resolving small, homogenously connected architectures, we modeled this local micro-scale neural population using an Erdős-Rényi topology ($N=10$, $p=0.15$), with a mixture of excitatory and inhibitory connections. This explicitly tests the framework's ability to operate on the types of sparse random graphs commonly utilized in computational neuroscience. The network was simulated under a transient regime, and the inference engine was applied to recover the underlying network topology.
    In panel b is reported the results of the inference engine. In b.1 we report the true network weights, in b.2 the probability for the presence of an edge, and in b.3 the expected inferred weights. The results show that the inference engine is able to recover the underlying network topology with high accuracy, as depicted also in the ROC curve in panel c, where the AUROC is reported to be $0.70$. The relation between the true and the inferred network weights is reported in panel d, where the Spearman correlation coefficient is reported to be $\rho = 0.70$, indicating a strong agreement between the true and inferred regulatory weights. Finally, in panel e we report the results of the perturbation analysis, where we evaluated the ability of the probabilistic inferred networks to predict the downstream effects of a node knockout, comparing the predictions from the MAP point-estimate and the ensemble of networks, showing a more robust predictive capability for the ensemble model ($\rho_{ens} = 0.47$), compared to the MAP-estimate ($\rho_{map}=0.31)$.}
    \label{fig:W-C_model}
\end{figure*}

While our framework was initially demonstrated on synthetic data generated from the exact same SDE equations used for inference, in real-world scenarios empirical data rarely adhere perfectly to an assumed mathematical model. To rigorously evaluate the algorithm's robustness against such generative model mismatch, we tested the inference engine on transient trajectories generated by a Wilson-Cowan neural rate model. 
The Wilson-Cowan equations actively deviate from our assumed SDE formulation by introducing hard non-linear refractory bounds. Despite these severe violations of the assumed internal physics, the inference engine successfully operated as a robust structural classifier, recovering the underlying topology with high accuracy ($AUPRC = 0.60$; Fig.~\ref{fig:W-C_model}). Furthermore, when simulating downstream perturbations on this inferred ensemble, the model accurately predicted the macroscopic systemic response, correctly identifying the direction and relative magnitude of the new steady-states. This success stems from the fact that, although the precise mathematical formulations differ, the qualitative phase-space flows of both models are sufficiently similar. The method does not rigidly require empirical data to perfectly adhere to a specific analytical equation; instead, it identifies the dominant structural constraints that shape these shared macroscopic dynamics. Full derivations and experimental details for this mismatch benchmark are provided in the Supplementary Materials.

\section{Conclusions}

Recovering the causal topology of stochastic, non-linear dynamical systems from sparse observations remains one of the most mathematically demanding inverse problems in complex systems. In this work, we introduced a fully Bayesian, ensemble-based framework that integrates a dynamical SDE likelihood with a non-parametric Minimum Description Length (MDL) structural prior via a Stochastic Block Model (SBM). By treating the negative log-likelihood of the trajectories as an Onsager-Machlup action functional, we established a rigorous theoretical foundation for path-weighting in continuous non-equilibrium state spaces.

Our results highlight four fundamental principles for robust network inference. First, we demonstrated that overparameterization of intrinsic node dynamics leads to severe structural degeneracy; globally constraining kinetic parameters is essential to force the inference engine to recover genuine network topology rather than absorbing variance through localized kinetic overfitting. Second, we showed that inference algorithms must exploit non-equilibrium, transient trajectories. Steady-state data lacks the necessary causal information, whereas transients traverse steep quasi-potential gradients that decisively constrain the directional flow of the network. Third, we established that the algorithm scales robustly to larger networks; despite the quadratically expanding parameter space and plummeting baseline sparsity, the Bayesian framework successfully leverages proportional data budgets to maintain high structural resolution. Finally, we proved that relying on a single Maximum A Posteriori (MAP) point-estimate is highly brittle for downstream predictive modeling. By marginalizing over topological uncertainty, our posterior ensemble accurately forecasted the macroscopic systemic effects of network perturbations, capturing massive regulatory shifts that point-estimates entirely missed.

It is important to note that the very constraint making this inference possible---globally shared kinetic parameters---establishes a fundamental boundary condition. When applied to systems governed by heavily divergent, node-specific thermodynamic rates, this homogeneous constraint induces a model mismatch that degrades structural recovery. However, as demonstrated by our Wilson-Cowan benchmark, the method is remarkably robust to severe violations of the assumed mathematical equations---such as hard non-linear bounds---provided the underlying system's macroscopic phase-space flows remain qualitatively similar. Consequently, this approach is best suited for complex systems where the primary variation in dynamics is driven by topological structure rather than massive intrinsic kinetic heterogeneities. 

While this framework was benchmarked primarily on macroscopic population dynamics, its mathematical foundation is entirely general. Future work will focus on expanding the algorithmic scalability and incorporating pseudo-time trajectory ordering to infer \textit{in vivo} developmental networks from single-cell transcriptomics. Ultimately, bridging the gap between rigorous MDL structural priors and non-linear continuous dynamics, while embracing and propagating structural uncertainty, offers a principled path toward predictive and interpretable models across all domains of complex systems.

\section{Materials and methods}

\subsection{Synthetic network generation and SDE integration}
Ground truth networks of sizes $N \in \{10, 20, 50, 100\}$ were generated as directed scale-free graphs using the Price model ($m=1$) to mimic the presence of highly connected hub nodes. The continuous time-series trajectories were generated by numerically integrating the stochastic differential equation (SDE) model. The wild-type global kinetic parameters were set to a basal degradation rate $\gamma = 0.5$, a maximum coupling strength $V = 2.0$, and an intrinsic noise level $D = 0.02$. The integration was performed using the Euler-Maruyama scheme with a highly resolved internal time-step of $\delta t = 0.01$.

To construct the training datasets, the trajectories were sub-sampled to simulate experimental measurement intervals $\Delta t \in \{0.01, 0.03, 0.05, 0.1, 0.5, 1.0\}$. Furthermore, data pairs were extracted from two distinct temporal regimes to evaluate the impact of non-equilibrium dynamics: a highly transient phase (sampling window $t \in [0, 3]$) and a stationary phase (sampling window $t \in [10, 15]$). The number of sampled state-transition pairs $M_{pairs}$ was systematically varied between 100 and 1000. 

\subsection{Wilson-Cowan neural rate model simulation}
To test the robustness of our inference framework against model mismatch, we simulated continuous time-series data using the Wilson-Cowan neural rate model:

\begin{equation}
    \frac{dx_i}{dt} = -\frac{x_i}{\tau} + (1 - x_i)\sigma\left(\sum_{j=1}^{N} W_{ij} x_j + I_i\right) + \eta_i(t)
\end{equation}

where $\sigma(\cdot)$ is the standard sigmoid function acting as the neuronal activation curve, $\tau=2.0$ represents the decay time constant, $I_i=-1.5$ is the external baseline input, and $\eta_i(t)$ represents additive Gaussian noise with standard deviation $0.05$. We simulated $N=10$ nodes connected by an Erd\"os-R\'enyi topology ($p=0.15$) with mixed excitatory and inhibitory connections. The non-zero weights were sampled uniformly from $W_{ij} \sim U(0.5, 1.5)$ and assigned random signs. We generated 30 independent transient trajectories of 200 integration steps ($\delta t = 0.01$), which were subsequently subsampled by a factor of 10 to simulate experimental observations.

\subsection{Bayesian inference and MCMC sampling}
The non-parametric Minimum Description Length (MDL) prior and the Stochastic Block Model (SBM) degree-corrected components were implemented via the \textit{graph-tool} Python library. The inference algorithm was initialized with an empty graph supplemented by a $5\%$ density of random edges to prevent the Markov Chain Monte Carlo (MCMC) sampler from trapping in trivial empty-network local minima.

The structural topology was sampled via the $NN_{Descent}$ block-coordinate sweeps, while the continuous global kinetic parameters $\theta = \{\gamma, V, D\}$ were optimized deterministically conditioned on the topology using the Nelder-Mead simplex algorithm. The inference pipeline consisted of 50 burn-in sweeps to allow the algorithm to converge near the global optimum, followed by 100 posterior sampling sweeps from which the probabilistic ensemble of networks was extracted.

\subsection{Evaluation metrics}
The probabilistic ensemble generated by the MCMC sampler was evaluated using a combination of topological and hierarchical metrics. The ensemble networks were marginalized to compute individual edge probabilities. These probabilities were used as continuous confidence scores to compute the Area Under the Precision-Recall Curve (AUPRC) against the binary ground-truth adjacency matrix. 

Due to inherent parameter non-identifiability, evaluating the absolute magnitudes of the inferred regulatory weights is less informative than evaluating the relative regulatory hierarchy. Therefore, the Spearman Rank correlation was used to compare the hierarchical ordering of the inferred dominant nodes against the true network. For perturbation analyses, the systemic shift in steady states was evaluated using a Kernel Density Estimation (KDE) log-likelihood metric, which measures the probability density assigned by the predicted ensemble to the true perturbed systemic state.

\subsection{Benchmarking baselines}
To benchmark against canonical inference paradigms, an extensive suite of nine widely used methods was employed on the same synthetic datasets: dynGENIE3 (non-parametric Random Forest inference adapted for time-series), Sparse ODE via Lasso (parametric linear regression with L1 regularization cross-validated for sparsity), Elastic Net VAR (parametric linear regression combining L1 and L2 regularization), Granger Causality (classical vector autoregression), PCMCI (information-theoretic causal discovery), Dynamic Bayesian Networks (discrete-time probabilistic structure learning), Gaussian Process ODEs (non-parametric Bayesian regression), PySINDy (data-driven sequential thresholded least squares using a polynomial library), and Neural ODEs (highly parameterized continuous-time deep learning). Hyperparameters for these baselines were selected according to standard default recommendations for biological time-series, with complete implementation details provided in the Supplementary Material.

\subsection{In silico perturbation experiments}
To simulate a node ablation experiment, the true and inferred networks were subjected to an \textit{in silico} perturbation where the target node's activation state and outgoing regulatory weights were permanently forced to zero. For the specific Wilson-Cowan mismatch experiment, we instead applied a localized over-expression ($\Delta I_{ko} = +5.0$) to a single target node. The networks were then allowed to relax to their new steady-state distribution. Predictive generalization was evaluated by comparing the continuous probability density of the perturbed steady states marginalized over the entire inferred posterior ensemble against the deterministic prediction of the single Maximum A Posteriori (MAP) point-estimate.

\vspace{0.2 cm}

\section*{Data and code accessibility}
All code used for the inference engine, MCMC sampling, and generation of the synthetic datasets is publicly available on GitHub at \url{https://github.com/CoMuNeLab/Onsager_Machlup_Inf}.

\section*{Competing interests}
The authors declare that they have no competing interests.

\section*{Funding}
F.Z. is supported by a Decreto Ministeriale n. 118 del 02/03/2023, M4C1 I. 4.1 from Piano Nazionale di Ripresa e Resilienza (PNRR). V.d.A. and M.D.D. acknowledge financial support from the EU funding within the MUR PNRR ``National Center for HPC, BIG DATA AND QUANTUM COMPUTING'' (Project no. CN00000013 CN1) and from the INFN grant ``LINCOLN''. MDD acknowledges partial support from the EU's Horizon Europe research and innovation programme under grant agreement No. 101186013 (VirHox) and from MUR funding within the FIS (DD n. 1219 31-07-2023) Project no. FIS00000158 (CUP C53C23000660001)

\section*{Acknowledgments}
The authors acknowledge the generous support of CloudVeneto for the computing infrastructure.

\vspace{0.2 cm}
\vspace{0.2 cm}

{\bf Author contributions} 
F.Z., M.D.D. and V.d.A. designed the study. F.Z. performed the experiments and analyzed the data. All authors wrote the manuscript and approved the final version.

\vspace{0.2 cm}

\renewcommand{\bibsection}{\section{References}}
\bibliographystyle{plainnat}
\bibliography{biblio}

\clearpage
\onecolumngrid
\appendix
\setcounter{section}{0}
\renewcommand{\thesection}{S\arabic{section}}
\setcounter{figure}{0}
\renewcommand{\thefigure}{S\arabic{figure}}
\setcounter{equation}{0}
\renewcommand{\theequation}{S\arabic{equation}}
\setcounter{table}{0}
\renewcommand{\thetable}{S\arabic{table}}

\begin{center}
\textbf{\large Supplementary Material: \\ Resolving Structural Uncertainty in Stochastic Dynamical Systems via Ensemble-Based Bayesian Inference}
\end{center}
\vspace{0.5cm}

\section{Likelihood Convexity}

To determine whether the likelihood is convex or not with respect to the interaction matrix weights, we compute the Hessian matrix and verify that it is semi-positive definite.

We focus on a single node $i$. Under the Euler-Maruyama discretization, the objective function to minimize (the negative log-likelihood) over $M$ time steps is the sum of squared residuals:

$$ \mathcal{L}(\mathbf{W}_i) = \frac{1}{4D \Delta t} \sum_{k=1}^M (e_{i,k})^2 $$

Where the residual error $e_{i,k}$ at time step $k$ is defined as:

$$ e_{i,k} = x_{i, k} - x_{i, k-1} - \Delta t \Big( -\gamma_i x_{i, k-1} + V_i \sigma(z_{i,k-1}) \Big) $$

And $z_{i,k-1} = \sum_{j} W_{ij} x_{j,k-1}$ is the total input to the sigmoid function.

The first derivative is:
$$ \frac{\partial \mathcal{L}}{\partial W_{ij}} = \frac{1}{2D \Delta t} \sum_{k=1}^M e_{i,k} \frac{\partial e_{i,k}}{\partial W_{ij}} $$

Taking the derivative of the residual $e_{i,k}$ with respect to $W_{ij}$:
$$ \frac{\partial e_{i,k}}{\partial W_{ij}} = - \Delta t \, V_i \, \sigma'(z_{i,k-1}) \, x_{j, k-1} $$

So the gradient is:
$$ \frac{\partial \mathcal{L}}{\partial W_{ij}} = - \frac{V_i}{2D} \sum_{k=1}^M e_{i,k} \, \sigma'(z_{i,k-1}) \, x_{j, k-1} $$

Taking the second derivative, we get the entries of the Hessian matrix:

$$ H_{i,j} = \frac{\partial^2 \mathcal{L}}{\partial W_{ij} \partial W_{il}} = \frac{1}{2D \Delta t} \sum_{k=1}^M \left[ \frac{\partial e_{i,k}}{\partial W_{ij}} \frac{\partial e_{i,k}}{\partial W_{il}} + e_{i,k} \frac{\partial^2 e_{i,k}}{\partial W_{ij} \partial W_{il}} \right] $$

Unpacking each term:
    $$ \frac{\partial e_{i,k}}{\partial W_{ij}} \frac{\partial e_{i,k}}{\partial W_{il}} = \Delta t^2 V_i^2 [\sigma'(z_{i,k-1})]^2 \, x_{j, k-1} \, x_{l, k-1} $$
    
    $$ e_{i,k} \frac{\partial^2 e_{i,k}}{\partial W_{ij} \partial W_{il}} = - \Delta t \, V_i \, e_{i,k} \, \sigma''(z_{i,k-1}) \, x_{j, k-1} \, x_{l, k-1} $$

Combining both terms and factoring out 
$x_{j, k-1} x_{l, k-1}$, the full Hessian matrix element is:

$$ H_{jl} = \frac{V_i}{2D} \sum_{k=1}^M x_{j, k-1} x_{l, k-1} \underbrace{\left[ \Delta t \, V_i \big(\sigma'(z_{i,k-1})\big)^2 - e_{i,k} \, \sigma''(z_{i,k-1}) \right]}_{c_k} $$

In matrix form, the Hessian is $\mathbf{H} = \mathbf{X}^T \mathbf{C} \mathbf{X}$, where $\mathbf{C}$ is a diagonal matrix containing the coefficients $c_k$. 

\begin{equation}
    c_k = \Delta t \, V_i \big(\sigma'(z_{i,k-1})\big)^2 - e_{i,k} \, \sigma''(z_{i,k-1})
    \label{eq::ck}
\end{equation}

For the likelihood $\mathcal{L}$ to be strictly convex, the Hessian $\mathbf{H}$ must be Positive Semi-Definite (PSD) everywhere. This requires that the coefficient $c_k \ge 0$ for all time steps $k$.
Looking at \ref{eq::ck}, we can notice that the first term is always positive, while the second one can be negative, given that the second derivative of the sigmoid changes sign with respect to the sign of the argument.

This fact does not allow to be sure that a simple gradient descent algorithm can safely converge to the global minimum of the likelihood function. 

\section{Parameter Indistinguishability}

\subsection{Equilibrium Scaling Degeneracy ($\gamma_i \sim V_i$)}

Near dynamic equilibrium or steady state, the drift vanishes ($\Delta x_i(t) / \Delta t \approx 0$), yielding the stationary expectation:

$$x_i^* = \frac{V_i}{\gamma_i} \sigma\left( \sum_j W_{ij} x_j^* \right)$$

If both degradation $\gamma_i$ and transcription scale $V_i$ are multiplied by a scale factor $c > 0$:

$$\gamma_i \mapsto c \gamma_i, \quad V_i \mapsto c V_i \implies \frac{cV_i}{c\gamma_i} = \frac{V_i}{\gamma_i}$$

Thus, the stationary expression $x_i^*$ is invariant to uniform scaling of $(\gamma_i, V_i)$. In data lacking strong non-equilibrium transients, MCMC proposals along the ridge $\gamma_i = c V_i$ produce identical steady-state predictions.

\subsection{Sigmoid Saturation Degeneracy ($|m_i| \gg 1$)}

The derivative of the drift term with respect to an edge weight $W_{ij}$ is:

$$\frac{\partial f_i}{\partial W_{ij}} = V_i \frac{\partial \sigma(m_i)}{\partial m_i} x_j = V_i \sigma(m_i)\left(1 - \sigma(m_i)\right) x_j$$

When total regulatory input $|m_i| = \left|\sum_j W_{ij} x_j\right|$ is large, the sigmoid saturates ($\sigma(m_i) \to 0$ or $1$). In this saturated regime, the gradient vanishes:

$$\lim_{|m_i| \to \infty} \sigma(m_i)(1 - \sigma(m_i)) = 0$$

Consequently, once an edge $W_{ij}$ drives node $i$ into saturation, further changes to $W_{ij}$ or addition of new edges $W_{ik}$ produce virtually zero change in log-likelihood ($\Delta S \approx 0$), causing edge misinference.

\subsection{Multiplicative Linear Indeterminacy ($V_i W_{ij} = \text{const}$)}

When regulatory inputs are small ($|m_i| \lesssim 1$), the sigmoid operates in its linear regime ($\sigma(m_i) \approx \frac{1}{2} + \frac{1}{4} m_i$). The drift simplifies to:

$$f_i(x) \approx -\gamma_i x_i + \frac{1}{2} V_i + \frac{1}{4} V_i \sum_j W_{ij} x_j$$

In this linear regime, scaling $V_i \mapsto \alpha V_i$ while inversely scaling edge weights $W_{ij} \mapsto \alpha^{-1} W_{ij}$ leaves the effective interaction strength $V_i W_{ij}$ unchanged, introducing a multiplicative degeneracy between node parameters and edge weights.

\section{Exact Analytical Transition Moments (Itô Isometry)}

The Euler-Maruyama discretization, while computationally simple, is mathematically guaranteed to become unstable at large measurement intervals ($\Delta t$), failing to capture the exponential relaxation of the degradation term and incorrectly modeling the variance. To overcome this, we derive the exact analytical transition moments under a zero-order hold assumption, where the regulatory input $m_i(t) = \sum W_{ij} x_j(t)$ is assumed constant over the interval $\Delta t$.

Under this assumption, the SDE for node $i$ reduces to a strictly linear CIR process:
\begin{equation}
    dx_t = (-\gamma x_t + b) dt + \sqrt{2D x_t} dW_t
\end{equation}
where $b = V \sigma(m_i(t))$. We define the steady-state asymptote as $\mu_\infty = b / \gamma$.

The exact expected mean $\mathbb{E}[x_{t+\Delta t} \mid x_t]$ is the solution to the deterministic ODE part:
\begin{equation}
    \mu(t+\Delta t) = \mu_\infty + (x_t - \mu_\infty)e^{-\gamma \Delta t}
\end{equation}

To find the exact accumulated variance, we apply Itô's Isometry. The variance of the integrated stochastic process over $[0, \Delta t]$ is:
\begin{equation}
    \text{Var}(x_{t+\Delta t} \mid x_t) = \mathbb{E}\left[ \int_0^{\Delta t} e^{-2\gamma(\Delta t - s)} 2D \cdot \mathbb{E}[x_s \mid x_t] ds \right]
\end{equation}
Substituting the exact expected mean $\mathbb{E}[x_s \mid x_t]$ into the integral and evaluating it yields the exact closed-form solution:
\begin{equation}
    \text{Var} = \frac{2D}{\gamma} (x_t - \mu_\infty) (e^{-\gamma \Delta t} - e^{-2\gamma \Delta t}) + \frac{D \mu_\infty}{\gamma} (1 - e^{-2\gamma \Delta t})
\end{equation}
This formula correctly collapses to the Euler-Maruyama variance $2D x_t \Delta t$ as $\Delta t \to 0$, and saturates to the finite steady-state variance $D \mu_\infty / \gamma$ as $\Delta t \to \infty$.

\section{Evaluation Metrics Formulation}

Because the inference produces a posterior ensemble rather than a single point-estimate, we evaluate the structural recovery using probabilistic metrics based on the marginal edge probability $p_{ij}$, which is the fraction of MCMC sweeps in which the edge $i \to j$ exists. The Area Under the Receiver Operating Characteristic (AUROC) and Precision-Recall Curve (AUPRC) were calculated by thresholding $p_{ij}$ directly, bypassing binary MAP predictions.

For the perturbation analysis, where the goal is to predict the downstream steady-state $x_j$ after a knockout or over-expression, we evaluate the ensemble's predictive power using a \textbf{Kernel Density Estimation (KDE) log-likelihood}. We construct a probability density function $f(x)$ from the ensemble of predicted steady-states using a Gaussian kernel, and evaluate the likelihood of the true systemic steady-state $x^{true}$ under this density:
\begin{equation}
    \text{LL}_{KDE} = \log f(x^{true}) = \log \left( \frac{1}{S} \sum_{s=1}^S \mathcal{N}(x^{true} \mid x^{(s)}, h^2) \right)
\end{equation}
where $S$ is the number of ensemble samples and $h$ is the bandwidth chosen via Scott's rule.

\section{Baseline Competitor Hyperparameters}

To ensure a fair and reproducible benchmarking study against standard inference paradigms, the baseline methods were executed using the following settings:

\begin{itemize}
    \item \textbf{dynGENIE3:} Executed using the default Random Forest implementation tailored for time-series data. The number of trees was set to 1000 to ensure stable variable importance metrics.
    \item \textbf{Sparse ODE via Lasso (Inferelator-style):} Implemented as a regularized linear regression problem predicting the finite-difference derivative $\Delta x_i / \Delta t$. The $L_1$ regularization penalty ($\alpha$) was optimized independently for each node using 5-fold cross-validation (LassoCV) to enforce sparsity.
    \item \textbf{PySINDy:} The Sequential Thresholded Least Squares (STLSQ) optimizer was used. To remain comparable with our SDE model's capacity, the feature library was restricted to first-order polynomials (linear combinations of the node states). The sparsity threshold was optimized via cross-validation.
    \item \textbf{Granger Causality:} Executed using a classical Vector Autoregression (VAR) model. The optimal maximum lag was selected using the Akaike Information Criterion (AIC). F-tests were performed to determine the statistical significance of predictive causality, and the resulting p-values ($- \log_{10}(p)$) were used as confidence scores for the AUPRC evaluation.
    \item \textbf{Elastic Net VAR:} Implemented as an extension to standard Lasso using Scikit-Learn's \texttt{ElasticNet}. It combined $L_1$ and $L_2$ penalties, fitting independent regression models mapping $\mathbf{X}(t)$ to $\frac{d\mathbf{X}}{dt}$. The resulting coefficients served directly as the inferred continuous causal interaction matrix $W$.
    \item \textbf{Gaussian Process ODEs (GP-ODEs):} Fit using Gaussian Process regressors with an Automatic Relevance Determination (ARD) Radial Basis Function (RBF) kernel. The inverse of the learned length scales for each input dimension was extracted as a proxy for the causal interaction strength.
    \item \textbf{Neural ODEs:} Implemented via PyTorch using a Multi-Layer Perceptron (MLP) with a single hidden layer to map $\mathbf{X}(t)$ to $\frac{d\mathbf{X}}{dt}$ via Mean Squared Error minimization over 200 epochs. To extract a linear causal skeleton, the Jacobian of the learned function $\frac{\partial f_i}{\partial X_j}$ was evaluated at the empirical mean of the training data.
    \item \textbf{PCMCI:} Implemented via the \texttt{tigramite} library. Simulated transient pairs were concatenated to form sequential transitions. The resulting p-value matrices for lagged links ($\tau = 1$) via Partial Correlation tests were converted to continuous confidence scores ($1 - p$).
    \item \textbf{Dynamic Bayesian Networks (DBNs):} Discretized the continuous states into equal-frequency bins and utilized \texttt{pgmpy}. Structure learning was performed across time-slices $t$ and $t+1$ using Hill Climb Search optimized via the Bayesian Information Criterion (BIC), blacklisting contemporaneous and backward-in-time edges.
\end{itemize}

\section{Erd\H{o}s-R\'enyi Grid Search Benchmark}

\begin{figure*}[h!]
    \centering
    \includegraphics[width=\textwidth]{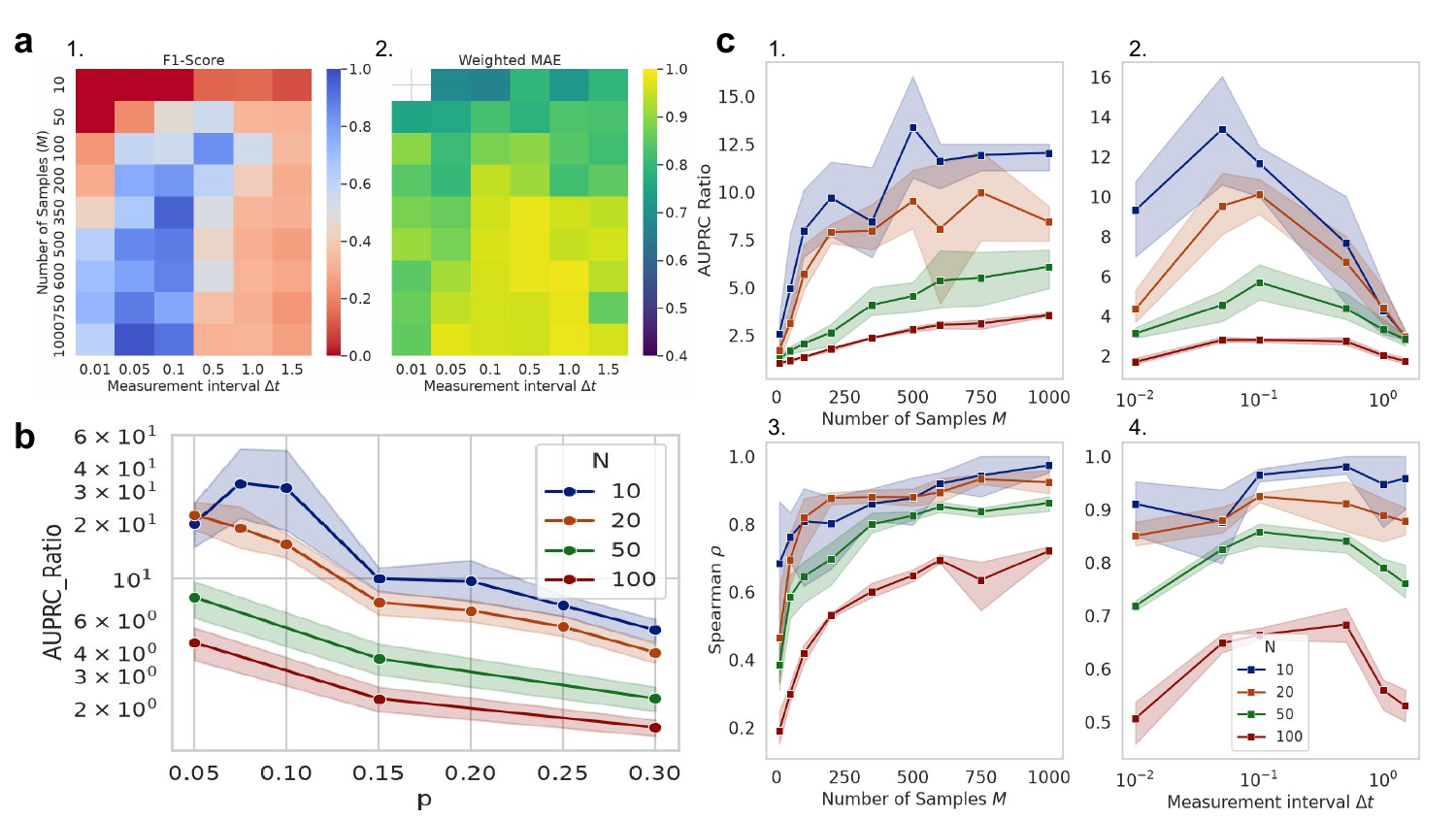}
    \caption{\textbf{Grid search performance on Erd\H{o}s-R\'enyi (ER) networks.} 
    \textbf{a.} Performance heatmaps as a function of the measurement interval $\Delta t$ and the number of sample pairs $M$ for an ER network of size $N=10$ ($p=0.15$) in the transient regime. 
    \textbf{a.1} reports the F1-Score, showing topological recovery accuracy, while \textbf{a.2} reports the Weighted Mean Absolute Error (MAE), quantifying the reconstruction accuracy of the regulatory weights.
    \textbf{b.} Scalability with edge density: Area Under the Precision-Recall Curve Ratio (AUPRC Ratio) as a function of the edge probability $p$ across network sizes $N \in \{10, 20, 50, 100\}$ at fixed sampling interval $\Delta t=0.05$. For small networks ($N \le 20$), intermediate densities ($p \approx 0.075\text{--}0.10$) maximize the performance relative to a random baseline, whereas for larger networks ($N=50, 100$), performance decreases with $p$ due to topological confounding from high average degrees ($k \approx pN$).
    \textbf{c.} Scaling of structural and parameter inference with sample size and temporal resolution for fixed edge probability $p=0.15$ in the transient regime across network sizes $N \in \{10, 20, 50, 100\}$.
    \textbf{c.1} and \textbf{c.2} show the AUPRC Ratio as a function of sample size $M$ (at fixed $\Delta t=0.05$) and measurement interval $\Delta t$ (at fixed $M=500$), respectively.
    \textbf{c.3} and \textbf{c.4} show the Spearman rank correlation coefficient $\rho$ of the inferred regulatory weights as a function of $M$ (at fixed $\Delta t=0.05$) and $\Delta t$ (at fixed $M=500$), respectively. Shaded bands represent standard deviations across independent realizations.}
    \label{fig:grid_search_er}
\end{figure*}

To ensure the scalability of our Bayesian framework was not strictly dependent on the presence of topological hubs, we replicated the comprehensive grid search on homogeneously random Erd\H{o}s-R\'enyi (ER) networks (Fig.~\ref{fig:grid_search_er}). For small networks ($N \le 20$), the ER topology enabled superior quantitative parameter inference (Spearman $\approx 0.86$, compared to $0.76$ for Scale-Free; Fig.~\ref{fig:grid_search_er}a,c.3). Lacking dominant hubs, ER networks distribute dynamical variance evenly, preventing the ``shadowing'' effect of master regulators and allowing the algorithm to precisely pinpoint individual edges. Within this regime, structural recovery peaked at intermediate edge probabilities ($p \approx 0.075 - 0.10$, AUPRC Ratio up to $33\times$ random; Fig.~\ref{fig:grid_search_er}b); extreme sparsity lacked sufficient causal coupling, while excessive density ($p \ge 0.30$) induced intractable topological confounding via indirect pathways. 

However, as dimensionality scaled ($N \to 100$), topological inference on ER networks degraded significantly faster than on Scale-Free baselines (Fig.~\ref{fig:grid_search_er}b,c). Because the average ER degree scales as $k \approx pN$, maintaining constant $p$ in large dimensions yields vastly denser networks (e.g., $k \approx 30$ for $N=100, p=0.30$), collapsing the AUPRC Ratio to $1.43$. Consequently, structural resolution in high dimensions strictly requires the absolute number of causal pathways per node to remain sparse ($p=0.05$). Unlike purely random ER networks, which become exceedingly difficult to untangle from homogeneous background noise at scale, real-world Scale-Free topologies provide the necessary dynamical hubs that act as structural anchors, allowing Bayesian inference to robustly scale to macroscopic complex systems.

\end{document}